\documentclass[pre,superscriptaddress,showpacs,twocolumn]{revtex4}

\usepackage{graphicx}%
\usepackage{dcolumn}
\usepackage{amsmath}

\usepackage[dvips]{epsfig}
\usepackage{amssymb}
\usepackage{array,longtable}
\usepackage{amscd}

\begin{document}



\newcommand{\be}{\begin{eqnarray}}
\newcommand{\ee}{\end{eqnarray}}
\newcommand{\bse}{\begin{subequations}}
\newcommand{\ese}{\end{subequations}}

\newcommand{\bs}{\boldsymbol}
\newcommand{\mbb}{\mathbb}
\newcommand{\mcal}{\mathcal}
\newcommand{\mfr}{\mathfrak}
\newcommand{\mrm}{\mathrm}

\newcommand{\ovl}{\overline}
\newcommand{\p}{\partial}
\newcommand{\f}{\frac}
\newcommand{\diff}{\mrm{d}}
\newcommand{\lan}{\left\langle}
\newcommand{\ran}{\right\rangle}

\newcommand{\ga}{\alpha}
\newcommand{\gb}{\beta}
\newcommand{\gc}{\gamma}
\newcommand{\Gd}{\Delta}
\newcommand{\gd}{\delta}
\newcommand{\Gc}{\Gamma}
\newcommand{\gl}{\lambda}
\newcommand{\gk}{\kappa}
\newcommand{\go}{\omega}
\newcommand{\Go}{\Omega}
\newcommand{\Gs}{\Sigma}
\newcommand{\gs}{\sigma}
\newcommand{\veps}{\varepsilon}
\newcommand{\eps}{\epsilon}
\newcommand{\Gt}{\Theta}

\newcommand{\sn}{\mrm{sn}}
\newcommand{\cn}{\mrm{cn}}
\newcommand{\dn}{\mrm{dn}}
\newcommand{\am}{\mrm{am}}
\newcommand{\sech}{\mrm{sech}}
\newcommand{\sign}{\mrm{sign}}

\newcommand{\csp}{\;,\qquad\qquad}
\newcommand{\fa}{\forall\;}

\newcommand{\N}{\mbb{N}}
\newcommand{\R}{\mbb{R}}
\newcommand{\D}{\mcal{D}}
\newcommand{\Nn}{\mcal{N}}
\newcommand{\V}{\mcal{V}}

\newcommand{\im}{\mrm{image}\;}
\newcommand{\num}{\mrm{\#}}


\title{Relativistic Brownian motion:
From a microscopic binary collision model to the Langevin equation}
\author{J\"orn Dunkel}
\email{joern.dunkel@physik.uni-augsburg.de}
\author{Peter H\"anggi}
\affiliation{Institut f\"ur Physik, Universit\"at Augsburg,
 Theoretische Physik I,  Universit\"atstra{\ss}e 1, D-86135 Augsburg, Germany}

\date{\today}

\begin{abstract}
The Langevin equation (LE) for the one-dimensional relativistic
Brownian motion is derived from a microscopic collision model. The model assumes  that a heavy point-like Brownian particle interacts with the lighter heat bath particles via elastic hard-core collisions. First, the commonly known, non-relativistic LE is deduced from this model, by taking into account the non-relativistic conservation laws for momentum and kinetic energy.
Subsequently, this procedure is generalized to the relativistic case.
There, it is found that the relativistic stochastic force is still $\gd$-correlated (white noise) but does \emph{no} longer correspond to a Gaussian
white noise process. Explicit results for the friction and momentum-space diffusion coefficients are presented and discussed.
\end{abstract}


\maketitle

\section{Introduction}
\label{introduction}
The theories of the non-relativistic Brownian motion and special
relativity were introduced more than 100 years
ago~\cite{Ei05c,EiSm,UhOr30,Ch43,WaUh45, HanJun95, HaMa05}. Since
then, they have become cornerstones for our understanding of a wide
range of physical processes~\cite{FrKr05,HaMaNo05,La05,ReHa02,BaReHa96}.
This fact notwithstanding, the unification of both concepts  poses a
theoretical challenge  still nowadays (classical references are
\cite{Sc61,Ha65,Du65,GuRu78,Bo79a,BY81}; recent contributions
include~\cite{MoVi95,Po97,DeMaRi97,KoLi00,Wo04,De04,Ko05,Zy05,Ra05,DuHa05a,DuHa05b,DuHa06,Fi06,Fa06};
potential applications in high-energy physics and astrophysics are
considered
in~\cite{AbGa04,AbGa05,2006PhRvC..73c4913V,2006hep.ph....1166V,DiDrSh06,MaFoSi06a}).
The relatively slow progress in this field can be attributed to the
severe difficulties that arise when one tries to describe $N$-body
systems in a relativistically consistent
manner~\cite{WhFe49,DaWi65,Ko78b,DuNaTr01,Le06}. Due to this reason, the derivation of
relativistic Langevin equations (LEs) from an underlying microscopic
model has remained an unsolved issue until now~\footnote{For the
non-relativistic Brownian motion, this problem was solved by
Bogolyubov~\cite{Bo45}, Magalinskii~\cite{Ma59}, Ford {\it
et al.}~\cite{FoKaMa65} and Zwanzig~\cite{Zw73}, who considered a
bath of harmonic oscillators.}. However, in the present paper we aim
to provide a solution to this problem.
\par
More precisely, by considering quasi-elastic, binary collisions
between the Brownian and heat bath particles~\cite{Pe91,TsPe00} we are
able to treat the heat bath in a fully relativistic manner without
having to account for the exact details of the relativistic $N$-body
interactions. As shown in Sec.~\ref{s:nonrelativistic}, for a
non-relativistic framework this approach yields the well-known
non-relativistic LE with Gaussian white noise as well as the correct
fluctuation-dissipation theorem (the Einstein-Sutherland
relation~\cite{HaMa05}). In Sec.~\ref{s:relativistic}, the method is
transferred to the relativistic case, leading to the main result of
this paper, the relativistic LE~\eqref{e:langevin_rel}. Remarkably,
the relativistic stochastic force is also  $\gd$-correlated (\lq
white\rq) but no longer of Gaussian  (or Wiener~\cite{Wi23}) type.
Compared with the non-relativistic Brownian motion, this is the most
important difference. Furthermore, we obtain explicit
representations of friction and (momentum)-diffusion coefficients in
terms of expectation values with respect to the heat bath
distribution (see also App.~\ref{a:solution}).

\section{Non-relativistic Brownian motions}
\label{s:nonrelativistic}

The objective of this section is to recover the well-known
non-relativistic LEs  from a simple microscopic collision model for
Brownian motions. As shown by several authors in the past~\cite{Bo45,Ma59,FoKaMa65,Zw73,Ha97}, non-relativistic LEs can
also be derived by considering a bath of harmonic oscillators
with canonical phase space distribution. Unfortunately, 
 it is problematic to transfer this approach to the
relativistic case, because any instantaneous linear (or nonlinear)
interactions between Brownian and heat particles would violate the basic
principles of special relativity~\cite{WhFe49,DaWi65}. To circumvent this problem, we
will pursue a different method here,  using only the (non-)relativistic
microscopic conservation laws for energy and momentum, respectively,
 known to hold for elastic point-like, binary collisions (contact interactions~\cite{DaWi65}). Conceptually, our
approach is related to that of Pechukas \cite{Pe91} and
Pechukas-Tsonchev~\cite{TsPe00}, who considered a similar model in
the context of non-relativistic quantum Brownian motion
\cite{HaIn05}. Analogous approaches are also known from unimolecular rate theory, see e.g. Sec.~V in \cite{HaTaBo90}.

\subsection{Microscopic model}
\label{microscopic-nr}

For the sake of simplicity only, we will restrict ourselves
throughout to the one-dimensional ($1d)$ case. Generalizations to
higher space dimensions are in principle straightforward, but
certain calculations will become much more cumbersome (cf. comments at the end of App.~\ref{a:solution}).
 To start out, consider the following situation in the inertial
laboratory frame $\Gs_0$: A large one-dimensional box volume $\V\equiv
[-L/2,L/2]$ contains an ideal non-relativistic gas, consisting of $N$ small
point-like particles with identical masses $m$. The gas particles --
referred to as \lq heat bath\rq\space hereafter -- surround a
Brownian particle of mass $M \gg m$.  Due to frequent elastic collisions with heat bath particles,
the Brownian particle performs stochastic motions.

\subsubsection{Heat bath}
\label{basic-nr}
The coordinates and momenta of the heat bath particles are denoted
by $x_r\in[-L/2,L/2]$ and  $p_r\in(-\infty;\infty)$, respectively,
where $r=1,\ldots,N$. As usual, we make the following simplifying
assumption concerning the heat bath: The probability density function (PDF)
of the heat bath particles is a spatially homogeneous Maxwell
distribution, i.e., at each time $t>0$, the PDF reads
\be\label{e:nonrelativistic_bath}
f_\mrm{b}^N(x_1,\ldots,p_N)
&=&
\left(\f{\gl}{L}\right)^N
\label{e:nonrelativistic_bath_a}
\prod_{r=1}^N \exp\biggl(-\f{p_r^2}{2m k T}\biggr),\qquad
\ee
where $k$ is the Boltzmann constant, $T$ the temperature,
and $\gl=(2\pi mkT)^{-1/2}$. Thus, it is implicitly assumed that:
\begin{itemize}
\item the heat bath particles are \emph{independently} and \emph{identically} distributed;
\item the collisions with the Brownian particle do not significantly alter the bath  distribution.
\end{itemize}
These assumptions are justified, if the collisions between the gas
particles rapidly reestablish a spatially homogeneous bath distribution.

\subsubsection{Kinematics of single collision events}
\label{s:nonrelativistic_kinematics}
The momentum and energy balance per (elastic) collision reads
\be\label{e:kinematics}
E+\eps=\hat{E}+\hat{\eps},\qquad
P+p=\hat{P}+\hat{p}.
\ee
Here and below, capital letters refer to the Brownian particle and
small letters to particles forming the heat bath; quantities without
(with) hat-symbols refer to the state before (after) the
collision. In the non-relativistic case, we have, e.g., before the collision
\be\label{e:nonrelativistic_definitions}
P=MV,\qquad p=mv,\qquad
E=\f{P^2}{2M},\qquad \eps=\f{p^2}{2m},\quad
\ee
where $v$ and $V$ denote the velocities. Taking into account both
conservation of momentum and (kinetic) energy, one finds that the
change $\Gd P\equiv\hat{P}-P$ of the Brownian particle's
momentum per single collision is given by
\be\label{e:p-tilde-nonrel}
\Gd P=\label{e:p-tilde-nonrel-a}
\f{-2m}{M+m}\,P+\f{2M}{M+m} \,p.
\ee

\subsection{Derivation of the Langevin equation}
\label{langevin-nonrel}
The total momentum change $\gd P$ of the Brownian particle within the time interval $\tau$ can be written as
\be\label{e:initial}
\gd P(t)\equiv P(t+\tau)-P(t)=\sum_{r=1}^N \Gd P_r\; I_r(t,\tau),
\ee
where $I_r(t,\tau)\in\{0,1\}$  is the indicator function for a collision
with the heat bath particle $r$ during the interval $[t,t+\tau]$; i.e.
 $I_r(t,\tau)=1$ if a collision has occurred, and, otherwise,
  $I_r(t,\tau)=0$. In the $1d$ case,  the collision indicator can be written explicitly as
\bse\label{e:indicator}
\be
I_r(t,\tau)
&=&\notag
\Gt(X-x_r)\;\Gt(x'_r-X')\;\Gt(v_r-V)+\\
&&\;
\Gt(x_r-X)\;\Gt(X'-x'_r)\;\Gt(V-v_r),\qquad
\label{e:indicator_1}
\ee
where $X=X(t),x_r=x_r(t)$, and
\be\label{e:indicator_0}
X'=X+V\tau,\qquad x'_r=x_r+v_r\tau
\ee
\ese
are the projected particle positions at time $t+\tau$.  The Heaviside-function is  defined by
\be\notag
\Gt(x)=\begin{cases}
0,& x< 0;\\
1/2,& x=0;\\
1, & x> 0.
\end{cases}
\ee 
The expectation of the collision indicator with respect to the bath distribution, denoted by $\lan I_r(t,\tau)\ran_\mrm{b}$, gives the
probability that the bath particle $r$ collides with the Brownian
particle between $t$ and $t+\tau$. As shown in
App.~\ref{a:solution}, in the limit $\tau\to 0$, one finds
\bse\label{e:Stokes} \be\label{e:Stokes-a} \lan
I_r(t,\tau)\ran_\mrm{b}= \tilde{C}(V)\; \f{\tau}{L}= C(P)\;
\f{\tau}{L}, 
\ee 
with function $C(P)=\tilde{C}(V(P))$ given by the
integral formula 
\be \tilde{C}(V)&\equiv&\notag \f{1}{2}
\int_{V}^{\infty}\!\!\!\diff v_r\;(v_r-V)\;
\tilde{f}_\mrm{b}^1(v_r)+\\
&&\;\label{e:rate-b}
\f{1}{2}
\int^{V}_{-\infty}\!\!\!\diff v_r\; (V-v_r)\;
\tilde{f}_\mrm{b}^1(v_r).
\ee
\ese
Here, $\tilde{f}_\mrm{b}^1(v_r)$ is the one-particle velocity PDF of a heat bath particle.
We anticipate that Eqs.~\eqref{e:indicator} and \eqref{e:Stokes} remain valid in
the relativistic case as well, but then one has to insert the relativistic bath distribution in~Eq.~\eqref{e:rate-b}.
\par
However, in order to recover from Eqs.~\eqref{e:initial}--\eqref{e:Stokes}
the well-known non-relativistic LE, we still have to make a number of simplifying assumptions:
\par(i)
The time interval $\tau$ is sufficiently small, so that $|\gd P/P|\ll 1$.
In particular, $\tau$ is supposed to be so small that there occurs at most
only one collision between the Brownian particle and a specific heat bath
particle $r$. One the other hand, the time interval $\tau$ should still
be large enough, so that the total number of collisions within $\tau$ is
larger than $1$. These requirements can be fulfilled simultaneously only if $m/M\ll 1$.
\par(ii)
Collisions occurring within $[t,t+\tau]$ can be viewed as independent events.
\par(iii)
Finally, we will (have to) assume that
\be
\lan [p_r\; I_r(t,\tau)]^j\ran_\mrm{b}
&=& \notag
\lan p_r^j \;I_r(t,\tau)\ran_\mrm{b}\\
&\simeq& \notag
\lan p_r^j \ran_\mrm{b}\;\lan I_r(t,\tau)\ran_\mrm{b}\\
&=& \lan p_r^j \ran_\mrm{b}\;C(P)\; \f{\tau}{L} \label{e:Stokes-b}
\ee for $j=1,2,\ldots$. Given the explicit representation of the
indicator function~\eqref{e:indicator_1}, it is in principle
straightforward to check  the quality of  the
approximation~\eqref{e:Stokes-b}, if a bath distribution has been
specified.
\par
As we shall see immediately, the assumptions (i)--(iii) are
necessary and sufficient for deriving the well-known
non-relativistic LE from Eqs.~\eqref{e:initial}--\eqref{e:Stokes}.
Upon inserting Eq.~\eqref{e:p-tilde-nonrel} into \eqref{e:initial}
and dividing by $\tau$ we find
 \be\notag \f{\gd P(t)}{\tau}&\simeq&
-\left[\f{1}{\tau}\sum_{r=1}^N\f{2m}{m+M} \;I_r(t,\tau)\right] \,P + \\
\label{e:initial_1}
&&\qquad\f{1}{\tau}\sum_{r=1}^N\f{2M}{M+m} \,p_r\;I_r(t,\tau).\qquad
\ee
The first term on the rhs. in Eq.~\eqref{e:initial_1} can be identified as
the \lq friction\rq\space term, whereas the second term represents \lq noise\rq.
On the rhs. of Eq.~\eqref{e:initial_1}, it was assumed that for each collision
occurring within $[t,t+\tau]$, the \lq initial\rq\space momentum of the Brownian particle is
approximately equal to some suitably chosen value $P(t')$ with $t'\in [t,t+\tau]$,
cf. the assumption (i) above and the discussion at the end of this section.
\par
The next step \textit{en route} to the conventional LE consists in replacing the
square bracket expression in Eq. \eqref{e:initial_1} by the \emph{averaged} friction coefficient
\bse\label{e:nu_nonrel}
\be\label{e:nu_definition}
\nu_0(P)&\equiv&
\f{1}{\tau}\sum_{r=1}^N\f{2m}{m+M} \, \lan I_r(t,\tau)\ran_\mrm{b}.
\ee
Since it was assumed that the heat bath particles are independently and
identically distributed, we can rewrite this as
\be\label{e:nu0}
\nu_0(P)&=&\f{N}{\tau}\f{2m}{m+M} \, \lan I_r(t,\tau)\ran_\mrm{b},
\ee
\ese
for some $r\in\{1,\ldots,N\}$. The coefficient $\nu_0$ can be interpreted
as an average collision rate weighted by some mass ratio. Inserting
Eq.~\eqref{e:Stokes-a} into Eq.~\eqref{e:nu0} yields
\bse\label{e:nu0_approx}
\be
\nu_0(P)&=&n_\mrm{b}\,\f{2m}{m+M} \, C(P),
\ee
where $n_\mrm{b}=N/L$ is the density of the bath particles. In the case of
the Maxwell distribution, we can evaluate the integral \eqref{e:rate-b}, and find
\be
C(P)&=&\notag
\left(\f{kT}{2\pi m}\right)^{1/2}
\exp\biggl[-\f{m}{2kT}\left(\f{P}{M}\right)^{2}\biggr]+\\
&&\qquad
\f{P}{2M}\;\mrm{erf}\biggl[\left(\f{m}{2kT}\right)^{1/2}\f{P}{M}\biggr].
\ee \ese In particular, setting (see App.~\ref{a:solution})
\be\label{e:stokes-nonrel} C(P)\approx C(0)=\left(\f{kT}{2\pi
m}\right)^{1/2} \ee corresponds to the commonly used \emph{Stokes
approximation}.
\par
It then remains to analyze the \lq noise force\rq
\be
\xi(P,t)\equiv
\f{1}{\tau} \sum_{r=1}^N\f{2M}{M+m} \,p_r \;I_r(t,\tau),
\ee
corresponding to the last term in Eq. \eqref{e:initial_1}. The momentum dependence of the noise enters through the implicit $P$-dependence of the collision indicator functions~$I_r(t,\tau)$. To keep subsequent formulae as compact as possible, we shall use the abbreviation $\xi(t)\equiv\xi(P,t)$ in the remainder. Then, averaging over the bath distribution $f_\mrm{b}^N$ and using Eqs.~\eqref{e:Stokes-b}, we
find for the mean value
\bse
\be
\lan \xi(t)\ran_\mrm{b} =0.
\ee
Furthermore, assuming mutual independence of the collisions, the correlation function is obtained as
\be
\lan  \xi(t)\, \xi(s)\ran_\mrm{b}
&=&\notag
\f{\gd_{ts}}{\tau^2}\left(\f{2M}{M+m}\right)^2
\sum_{r=1}^N \,\lan p_r^2\; I_r^2(t,\tau)\ran_\mrm{b}\\
&\overset{\eqref{e:Stokes-b}}{\simeq}&\notag
\f{\gd_{ts}}{\tau^2}\left(\f{2M}{M+m}\right)^2
\sum_{r=1}^N \,m kT\;\lan I_r(t,\tau)\ran_\mrm{b}\\
&\overset{\eqref{e:nu_definition}}{=}&\label{e:nonrel_correlation}
\f{\gd_{ts}}{\tau}\left(\f{2M^2}{M+m}\right)\;  \nu_0 kT,
\ee
\ese
with $\gd_{ts}\in\{0,1\}$ denoting the Kronecker-symbol. To
obtain the second line, we have used that $I_r^2(t,\tau)=I_r(t,\tau)$,
and the simplifying assumption~\eqref{e:Stokes-b} that $I_r(t,\tau)$ and
$p_r$ are (approximately) independent random variables with respect to the bath distribution.
\par
Similar to~Eq.~\eqref{e:nonrel_correlation}, also the higher correlation
functions are determined by the corresponding moments of the Gaussian marginal
bath distribution~\eqref{e:nonrelativistic_bath_a}.  Thus, under the above
assumptions (i)--(iii), the non-relativistic stochastic force $\xi(t)$ corresponds
to Gaussian white noise (or a Wiener process~\cite{Wi23}, respectively).
\par
Finally, by substituting $\nu_0$ from Eq~\eqref{e:nu0_approx} for
the square bracket expression in Eq.~\eqref{e:initial_1} and
formally letting $\tau\to 0$, we recover from
Eq.~\eqref{e:initial_1} the well-known non-relativistic LE~\cite{
HaTo82,Ha97,VK03} \bse\label{e:langevin_nr}
\be\label{e:langevin_nr-a} 
\dot P=-\nu_0(P)\, P+\xi(t), 
\ee 
where $\xi(t)\equiv\xi(P,t)$ is a momentum-dependent Gaussian white noise force, characterized by \be
\lan \xi(t)\ran&=&0,\\
\lan \xi(t)\,\xi(s)\ran
&=& \label{e:langevin_nr-c} 
2\,D_0(P)\;\gd(t-s),
\ee
with (momentum-space) diffusion coefficient
\be\label{e:FDT}
D_0(P)=\f{M^2}{M+m}\;\nu_0(P)\, kT.
\ee
\ese
To obtain Eq.~\eqref{e:langevin_nr-c}, we used that $\gd_{st}/\tau\to \gd(t-s)$ for $\tau\to 0$, where $\gd(t-s)$ is the Dirac-function.
\par
In the limit $m/M\to 0$, Eq.~\eqref{e:FDT} reduces to the standard
\emph{fluctuation-dissipation theorem} $D_0=M\nu_0 kT$
\cite{HaMa05,Be67}. However, $\nu_0$ and $D_0$ are constants only if
one adopts the Stokes approximation~\eqref{e:stokes-nonrel},
cf.~App.~\ref{a:solution}. If one goes beyond the Stokes
approximation, then the noise in Eqs.~\eqref{e:langevin_nr} becomes
multiplicative with respect to $P$, and, therefore,
Eqs.~\eqref{e:langevin_nr} must  be complemented by a discretization
rule in this
case~\cite{Ito44,Ito51,Fisk63,St64,St66,Han78,Han80,HaTo82,Kl94}. As
discussed in ~\cite{Han78,Han80,HaTo82,Kl94}, only for the
post-point discretization rule, corresponding to the choice 
$\nu_0(P)=\nu_0(P(t+\tau))$ and $D_0(P)=D_0(P(t+\tau))$ on the rhs.
of Eq.~\eqref{e:langevin_nr-a}, one recovers the Maxwellian PDF 
\be
\Phi_\infty(P)= \left(\f{1}{2\pi MkT}\right)^{1/2}
\exp\biggl(-\f{P^2}{2M k T}\biggr) 
\ee 
as the stationary momentum
distribution of the Brownian particle in the limit $t\to \infty$
(assuming that \mbox{$m/M\to 0$}).

\section{Relativistic Brownian motions}
\label{s:relativistic}
We shall now apply an analogous reasoning to obtain a relativistic LE.
For this purpose we consider an inertial (laboratory) frame $\Gs_0$ with
time coordinate $t$, as e.g. measured by an atomic clock resting in $\Gs_0$.

\subsection{Microscopic model}

The basic constituents of the microscopic model are the same as
those outlined in Sec.~\ref{microscopic-nr}, but in addition we now
have to consider a relativistic heat bath distribution and must
consistently take into account the relativistic collision
kinematics.

\subsubsection{Relativistic heat bath}

In the relativistic case, we postulate analogous to
Eq.~\eqref{e:nonrelativistic_bath} that, with respect to $\Gs_0$,
the heat bath distribution is stationary, spatially homogeneous, and
independent, so that the PDF can be written in the product form
\bse\label{e:relativistic_bath} 
\be\label{e:relativistic_bath_a}
f_\mrm{b}^N(x_1,\ldots,p_N) &=& \label{e:rel_pdf_ansatz} L^{-N}
\prod_{r=1}^N f_\mrm{b}^1(p_r).\qquad 
\ee 
As marginal one-particle
momentum PDFs, we will now consider the $\eta$-generalized
J\"uttner-Maxwell distributions~\cite{Ju11,DuHa06}, reading:
\be\label{e:Juettner} 
\label{e:Juettner-1d} 
f_\mrm{b}^1(p)
=\f{\mcal{N}_\eta}{\eps(p)^\eta} \exp\biggl[-\f{\eps(p)}{k
T}\biggr],\qquad \eta\ge 0, 
\ee 
where $p\in (-\infty,+\infty)$, and
$\eps(p)$ denotes the relativistic kinetic energy of a heat bath
particle. The normalization constant $\mcal{N}_\eta$ is determined
by the condition
\be 
1=\int_{-\infty}^\infty \diff
p\;f_\mrm{b}^1(p). 
\ee 
\ese 
For $\eta=0$, Eq.~\eqref{e:Juettner-1d}
reduces to the standard J\"uttner-Maxwell distribution~\cite{Ju11}.
On the other hand, as discussed recently~\cite{DuHa06,MaFoSi06a}, the PDF with $\eta=1$ appears to be conserved in relativistic elastic binary collisions.
In general, however, the arguments and results presented below
remain valid for arbitrary one-particle momentum
PDFs~$f^1_\mrm{b}(p)$, i.e., also for momentum distributions other
than the $\eta$-generalized J\"uttner PDFs~\eqref{e:Juettner}.

\subsubsection{Relativistic collision kinematics}

Using natural units such that $c=1$,  relativistic kinetic energy, momentum and velocity are related by
\bse\label{e:rel_kinematics}
\begin{align}
p&=mv\,\gc(v),&\qquad
\eps(p)=&\left(m^2+p^2\right)^{1/2},\\
P&=MV\,\gc(V),&\qquad
E(P)=&\left(M^2+P^2\right)^{1/2},
\end{align}
\ese
where $\gc(v)\equiv\left(1-v^2\right)^{-1/2}$. As before, capital letters
refer to the Brownian particle. Inserting Eqs.~\eqref{e:rel_kinematics} into
the conservation laws ~\eqref{e:kinematics}, and solving for $\hat{P}$, one finds~\cite{DuHa06}
\be
\hat{P}&=&\f{2u\, E-(1+u^2)\,P}{1-u^2},
\ee
where
\be\label{a-e:u}
u(p,P)=\f{P+p}{E+\eps}
\ee
is the center-of-mass velocity. Hence, the momentum change $\Gd P=\hat{P}-P$ of
the Brownian particle in a single collision is given by
\be\label{e:rel_delta}
\Gd P=-\f{2}{1-u^2}\; \f{\eps}{E+\eps} \;P+\f{2}{1-u^2}\; \f{E}{E+\eps}\; p.\ee
In the non-relativistic limit case, where $u^2\ll 1$, $E\simeq M$ and  $\eps\simeq m$,
this reduces to Eq.~\eqref{e:p-tilde-nonrel}.

\subsection{Derivation of the Langevin equation}
\label{langevin-rel}
Inserting Eq.~\eqref{e:rel_delta} into Eq.~\eqref{e:initial}, one obtains the
relativistic analogon of Eq.~\eqref{e:initial_1} as
\be\notag
\f{\gd P(t)}{\tau}&\simeq&
-\left[\f{1}{\tau}\sum_{r=1}^N \f{2}{1-{u}_r^2}\; \f{\eps_r}{E+\eps_r}\;I_r(t,\tau)\right] P + \\
&&\qquad
\f{1}{\tau}\sum_{r=1}^N
\f{2}{1-{u}_r^2}\; \f{E}{E+\eps_r} \,p_r\;I_r(t,\tau),
\label{e:initial_1_rel}
\ee
where ${u}_r\equiv u(p_r,P)$ and $\eps_r\equiv \eps(p_r)$. Formally, the collision
indicator $I_r(t,\tau)$ is still determined by Eqs.~\eqref{e:indicator} and \eqref{e:Stokes},
but differences arise due to the fact that we have to use $V=P/(M^2+P^2)^{1/2}$ and a
relativistic bath distribution now.
\par
\begin{figure}[h]
\center \epsfig{file=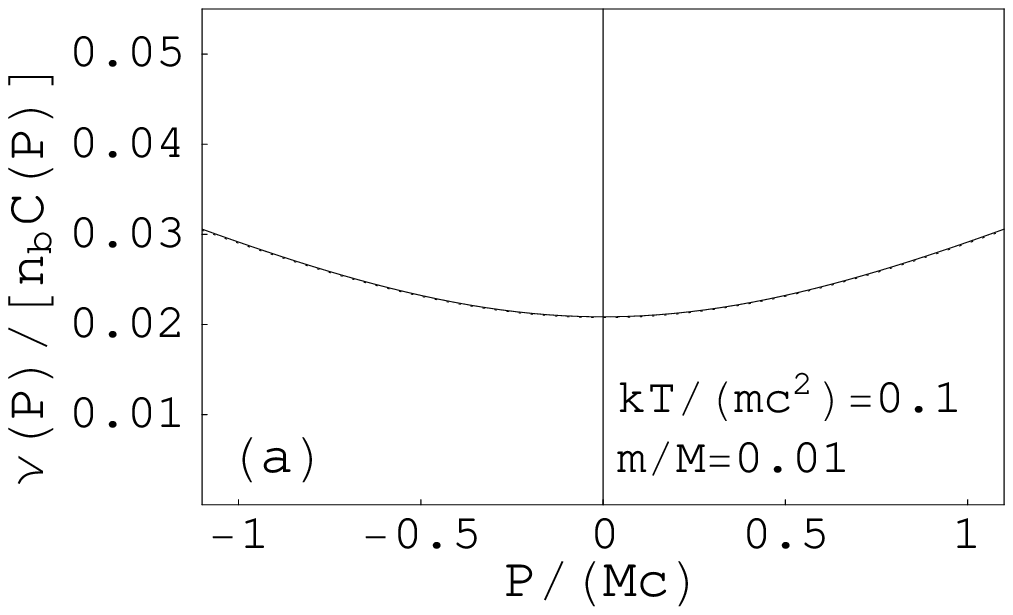 ,height=4.8cm, angle=0}
\epsfig{file=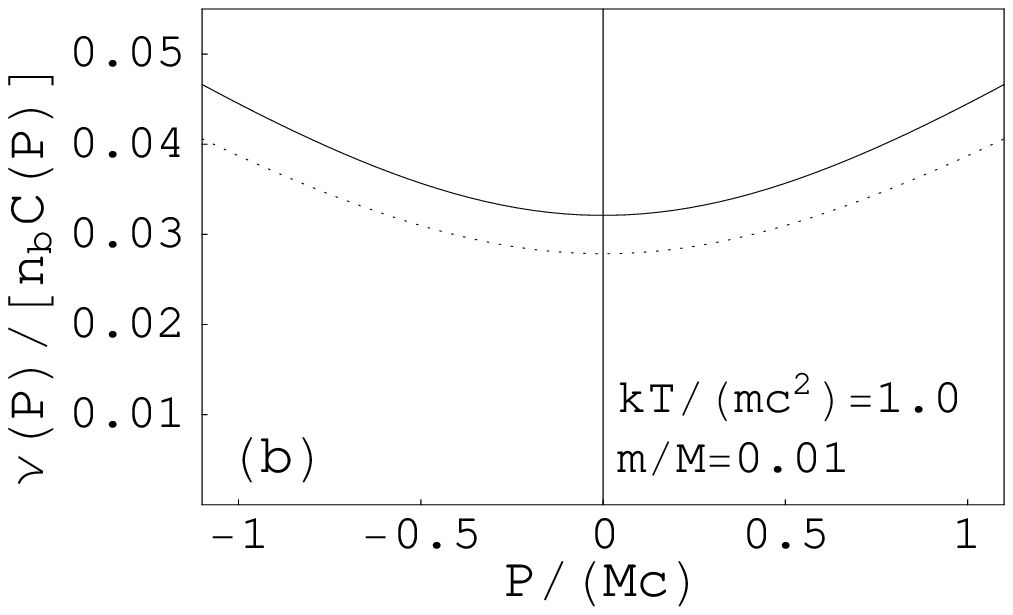 ,height=4.8cm, angle=0}
\caption{\label{fig01} The momentum-dependent, relativistic friction
coefficient $\nu(P)$, divided by the total mean collision rate,
$\nu(P)/[n_\mrm{b}C(P)]$, as calculated  numerically for two
different heat bath distributions $f_\mrm{b}^1(p)$ and two different
bath temperatures, is depicted versus the scaled momentum $P$. The
solid lines refer to the standard J\"uttner distribution with
$\eta=0$, and the dotted lines to $\eta=1$ in
Eq.~\eqref{e:Juettner-1d}. (a) \emph{Weakly relativistic heat bath.}
In the limit $kT\ll mc^2$ the bath
distributions~\eqref{e:Juettner-1d} approach a Maxwellian, and
therefore the results for different $\eta$ practically coincide. In
particular, for $P=0$ the non-relativistic result is recovered. (b)
\emph{Strongly relativistic heat bath.} The friction coefficient
increases with the temperature of the heat bath.}
\end{figure}
\begin{figure}[h] 
\center \epsfig{file=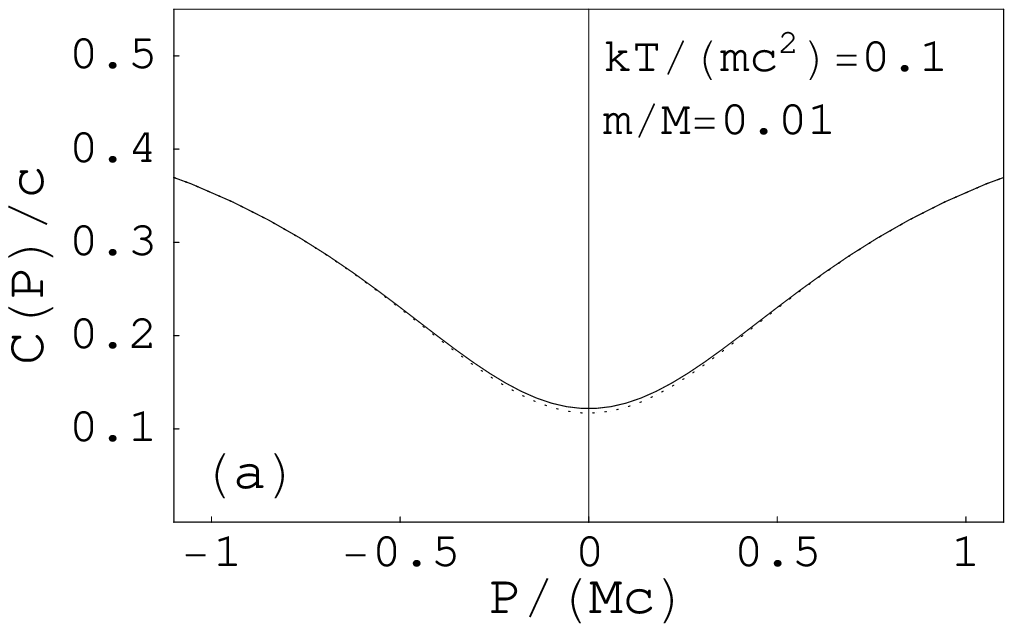 ,height=4.8cm, angle=0}
\epsfig{file=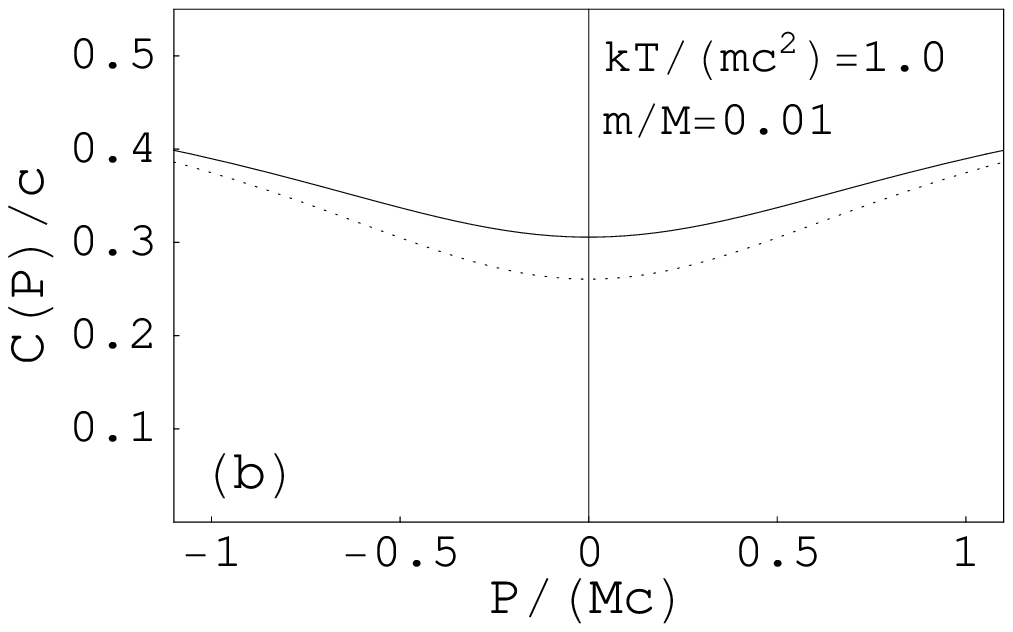 ,height=4.8cm, angle=0}
\caption{\label{fig02} Relativistic one-particle collision
coefficient $C(P)=\lan I_r(t,\tau)\ran_\mrm{b}L/\tau$, numerically
calculated for the same parameters as in Fig.~\ref{fig01}. The solid
lines refer to a standard J\"uttner distribution with $\eta=0$, and
the dotted lines to $\eta=1$ in Eq.~\eqref{e:Juettner-1d}. (a)
\emph{Weakly relativistic heat bath.}  At small temperatures, the
zero-value $C(0)$ is approximately equal to the non-relativistic
Stokes value $\sqrt{kT/(2\pi m)}$. (b) \emph{Strongly relativistic
heat bath.} For $|P|\to\infty$ the coefficient $C(P)$ converges to
$1/2$.}
\end{figure}

Analogous to the non-relativistic case, we can identify the first term on the rhs.
of Eq.~\eqref{e:initial_1_rel} as friction, and introduce an averaged friction coefficient by
\be
\nu(P)&\equiv&\notag
\f{1}{\tau}\sum_{r=1}^N \lan\f{2}{1-u_r^2}\;
\f{\eps_r}{E+\eps_r}\;I_r(t,\tau)\ran_\mrm{b}\\
&=&
\f{N}{\tau}\; \lan\f{2}{1-u_r^2}\; \f{\eps_r}{E+\eps_r}\;
I_r(t,\tau)\ran_\mrm{b},
\ee
for some $r\in\{1,\ldots,N\}$. Next, applying a product approximation
similar to~\eqref{e:Stokes-b}, we obtain
\be
\nu(P)&\simeq&\notag
\f{N}{\tau}\lan\f{2}{1-{u}_r^2}\; \f{\eps_r}{E+\eps_r}\ran_\mrm{b}\;
\lan I_r(t,\tau)\ran_\mrm{b}\\
&\overset{\eqref{e:Stokes}}{=}&\label{e:nu_rel}
n_\mrm{b}\; C(P)\lan\f{2}{1-{u}_r^2}\; \f{\eps_r}{E+\eps_r}\ran_\mrm{b},
\ee
where $n_\mrm{b}=N/L$ is the density of the heat bath particles, and
$C(P)$ is determined by Eq.~\eqref{e:rate-b}.  Figure~\ref{fig01}
shows the $P$-dependence of $\nu(P)/[n_\mrm{b}C(P)]$ for the bath
distributions from Eq.~\eqref{e:Juettner}. This momentum dependence
is induced by the appearance of  ${u}_r=u(p_r,P)$ and $E=E(P)$ in the
expectation value on the rhs. of Eq.~\eqref{e:nu_rel}. Furthermore, the
shape of the one-particle collision coefficient $C(P)=\lan I_r(t,\tau)\ran_\mrm{b}L/\tau$
is depicted in Fig.~\ref{fig02}. As one would intuitively expect, the
friction coefficient grows with the temperature $T$ of the heat bath
(at constant $P$) as well as with the absolute momentum of the Brownian particle (at constant $T$). In the non-relativistic limit case, where $u^2\ll 1$, $E\simeq M$ and  $\eps\simeq m$, the relativistic friction coefficient $\nu(P)$ from Eq.~\eqref{e:nu_rel} reduces to the non-relativistic result $\nu_0(P)$ from Eq.~\eqref{e:nu0_approx}.
\par
At this point, it might be worthwhile to emphasize once again
that \emph{product approximations of the form}
\be\label{e:product_approx}
\lan G(x_r,p_r)\; I_r(t,\tau)\ran_\mrm{b}\simeq
\lan G(x_r,p_r)\ran_\mrm{b}\; \lan I_r(t,\tau)\ran_\mrm{b},
\ee
as employed in Eq.~\eqref{e:Stokes-b} and also in the first
line of Eq.~\eqref{e:nu_rel}, \emph{can in principle be
omitted by using the explicit representation~\eqref{e:indicator}
of the collision indicator and the Eqs.~\eqref{a-e:mean} of the Appendix};
if one opts to avoid such approximations then the accuracy of the Langevin
model increases (note that this statement applies to the non-relativistic case, too).
However, in the following we shall continue to use Eq.~\eqref{e:product_approx}
in order to obtain a relativistic LE that is on an equal footing with the
non-relativistic LE~\eqref{e:langevin_nr}.
\par
\begin{figure}[h]
\center
\epsfig{file=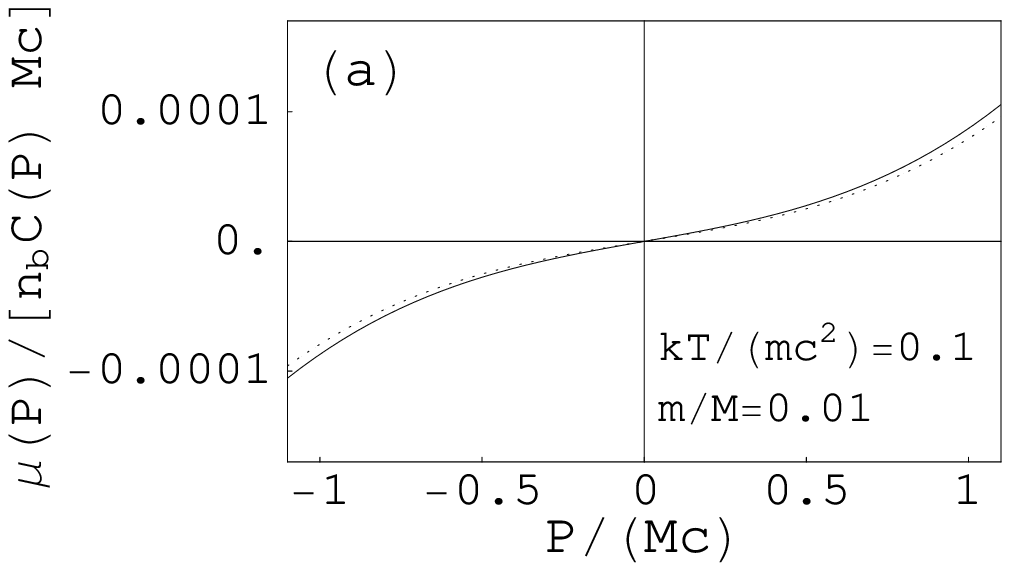 ,height=4.8cm, angle=0}
\epsfig{file=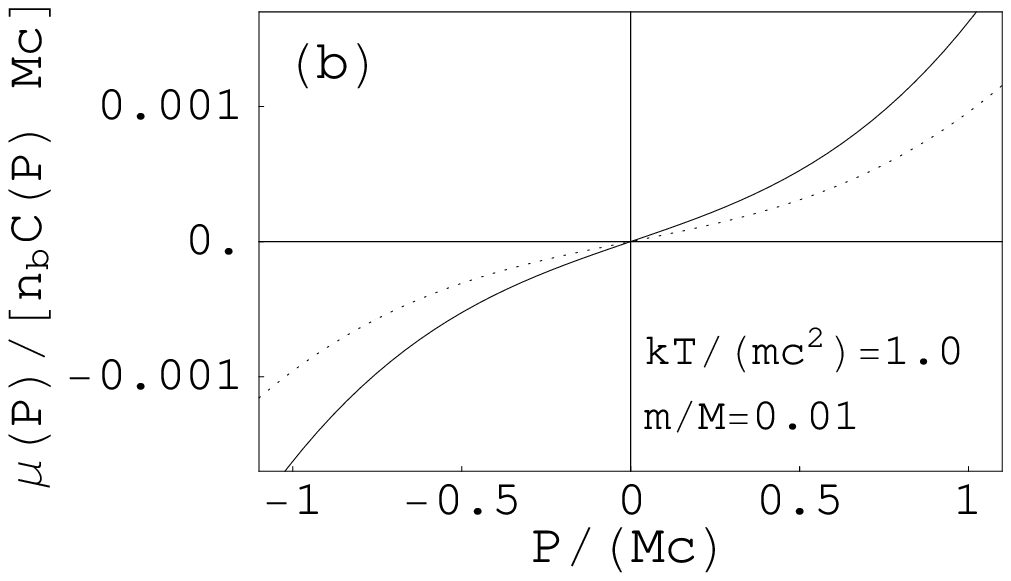 ,height=4.8cm, angle=0}
\caption{\label{fig03}
Mean value of the relativistic stochastic force, \mbox{$\mu(P)
\equiv\lan \chi(t)\ran_\mrm{b}$}, calculated  numerically for
two different heat bath distributions $f_\mrm{b}^1(p)$ and two
different bath temperatures. Solid lines refer to a
standard J\"uttner distribution with $\eta=0$,
and dotted lines to $\eta=1$ in Eq.~\eqref{e:Juettner-1d}.}
\end{figure}

For this purpose, we interpret the second term on the rhs. of
Eq.~\eqref{e:initial_1_rel} as \lq noise\rq, defining
\be
\chi(P,t)\equiv\f{1}{\tau}\sum_{r=1}^N
\f{2}{1-u_r^2}\; \f{E}{E+\eps_r} \,p_r\;I_r(t,\tau).
\ee
Averaging over the bath distribution $f_\mrm{b}^N$, one finds for the mean value
\be
\mu(P)&\equiv&\notag
\lan \chi(P,t)\ran_\mrm{b}\\
&=&\notag
\f{N}{\tau}
\lan \f{2}{1-u_r^2}\; \f{E}{E+\eps_r} \,p_r\;
I_r(t,\tau) \ran_\mrm{b}\\
&\overset{\eqref{e:product_approx}}{\simeq}&\notag
\f{N}{\tau}
\lan \f{2}{1-u_r^2}\; \f{E}{E+\eps_r} \,p_r\ran_\mrm{b}
\lan I_r(t,\tau) \ran_\mrm{b}\\
&\simeq&\label{e:mu}
n_\mrm{b}\;C(P)\;
\lan \f{2}{1-u_r^2}\; \f{E}{E+\eps_r} \,p_r\ran_\mrm{b}.
\ee
In contrast to the non-relativistic case, the mean value $\mu$ of the
relativistic Langevin force $\chi(t)\equiv\chi(P,t)$ depends on the momentum $P$ of
the Brownian particle. This can be attributed to the appearance
of \mbox{$u_r^2=(P+p_r)^2/(E+\eps_r)^2$} in Eq.~\eqref{e:mu}.
As shown in Fig.~\ref{fig03}, the quantity $\mu(P)/[n_\mrm{b}C(P)]$
is positive for $P>0$ and negative for $P<0$. Thus, on average,
the relativistic stochastic force tends to accelerate particles
in the direction of their motion, but this effect is compensated
by the increase of the friction coefficient $\nu(P)$ at high values of $P$, cf. Fig.~\ref{fig01}.
\par
Let us next take a closer look at the covariance function
\be\label{e:rel_correlation}
\gs_{ts}&\equiv&
\lan\, \left[\chi(t)-\lan \chi(t)\ran_\mrm{b}\right]\,
\left[\chi(s)-\lan \chi(s)\ran_\mrm{b}\right]\,\ran_\mrm{b}.
\ee
In the non-relativistic case, the stochastic force possesses a vanishing
mean value, $\lan \xi(t)\ran_\mrm{b}=0$.  According to Eq.~\eqref{e:mu},
this is no longer the case  for the relativistic noise force, $\lan \chi(t)\ran_\mrm{b}\ne 0$.
In order to explicitly calculate $\gs_{ts}$ for the relativistic case, it is convenient to introduce the abbreviation
$$   
\gk_r= \f{2}{1-u_r^2}\; \f{E}{E+\eps_r}\; p_r.
$$
Assuming, as before, that collisions can be viewed as independent
events, the correlation function~\eqref{e:rel_correlation}
vanishes at non-equal times $t\ne s$, and we thus find
\be
\gs_{ts}
&\overset{\eqref{e:mu}}{\simeq} &\notag
\gd_{ts}
\lan 
\left[\f{1}{\tau}\sum_{r=1}^N \gk_r I_r(t,\tau)\right]^2 -\mu^2(P)
\ran_\mrm{b}
\\
&\overset{\eqref{e:product_approx}}{\simeq}&\notag
\f{\gd_{ts}}{\tau^2} \biggl\{
\sum_{r=1}^N  \lan\gk_r^2 \ran_\mrm{b}\lan I_r(t,\tau)
\ran_\mrm{b}-\tau^2 \mu^2(P)
\\&&\notag\;
\sum_{r=1}^N \sum_{j\ne r}^N
\lan\gk_r \ran_\mrm{b}\lan I_r(t,\tau)\ran_\mrm{b}
\lan\gk_j \ran_\mrm{b}\lan I_j(t,\tau)\ran_\mrm{b}\biggr\}\\
&=&\notag
\f{\gd_{ts}}{\tau^2} \sum_{r=1}^N  \biggl\{
\lan\gk_r^2 \ran_\mrm{b}\lan I_r(t,\tau)
\ran_\mrm{b} -\f{\tau^2}{N} \mu^2(P)+\\
&&\notag\quad
\lan I_r(t,\tau)\ran_\mrm{b}\lan\gk_r \ran_\mrm{b}
\sum_{j\ne r}^N
\lan\gk_j \ran_\mrm{b}\lan I_j(t,\tau)\ran_\mrm{b}\biggr\}\\
&\overset{\eqref{e:mu}}{=}&\notag
\f{\gd_{ts}}{\tau^2} \sum_{r=1}^N  \biggl\{
\lan\gk_r^2 \ran_\mrm{b}\lan I_r(t,\tau)
\ran_\mrm{b}-\f{\tau^2}{N} \mu^2(P)+\\
&&\qquad\quad
\f{\tau}{N}\,\mu(P)
\sum_{j\ne r}^N \,\f{\tau}{N}\mu(P)
\biggr\}.\label{e:intermediate_results}
\ee
From this, we obtain
\be
\gs_{ts}
&{\simeq}&\notag
\f{\gd_{ts}}{\tau}\, \f{N}{\tau}
\lan\gk_r^2 \ran_\mrm{b}\lan I_r(t,\tau)
\ran_\mrm{b}-\f{\gd_{ts}}{N}\;\mu^2(P)\\
&\overset{\eqref{e:Stokes-a}}{=}&
\f{\gd_{ts}}{\tau}\,n_\mrm{b}\,C(P)\;
\lan\gk_r^2\ran_\mrm{b}-\f{\gd_{ts}}{N}\;\mu^2(P).
\notag
\\
\ee
The last term vanishes, if we consider the thermodynamic limit (TDL)
of an infinite heat bath, i.e., $N,L\to\infty$ such that $n_\mrm{b}=N/L=$constant.
Thus, reinserting the explicit expression for $\gk_r$, we obtain in this limit
\be
\gs_{ts}
\to\f{\gd_{ts}}{\tau}\,n_\mrm{b}\;C(P)
\lan\left(\f{2}{1-u_r^2}\; \f{E}{E+\eps_r}\; p_r\right)^2\ran_\mrm{b}.
\ee
In principle, any higher correlation function can be calculated in the same manner.
It is also evident that the noise force is \emph{non-Gaussian},
because the relativistic bath distribution $f_\mrm{b}(p_r)$ that
determines the averages $\lan\;\cdot\;\ran_\mrm{b}$ -- and, thus,
the noise correlations -- is  non-Gaussian.
\par
\begin{figure}[h]
\center \epsfig{file=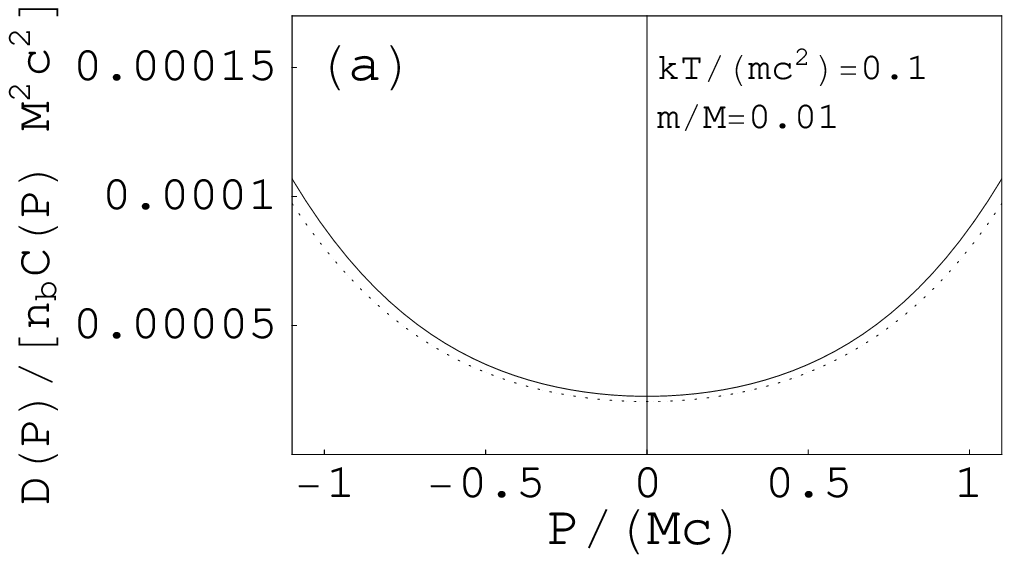 ,height=4.6cm, angle=0}
\epsfig{file=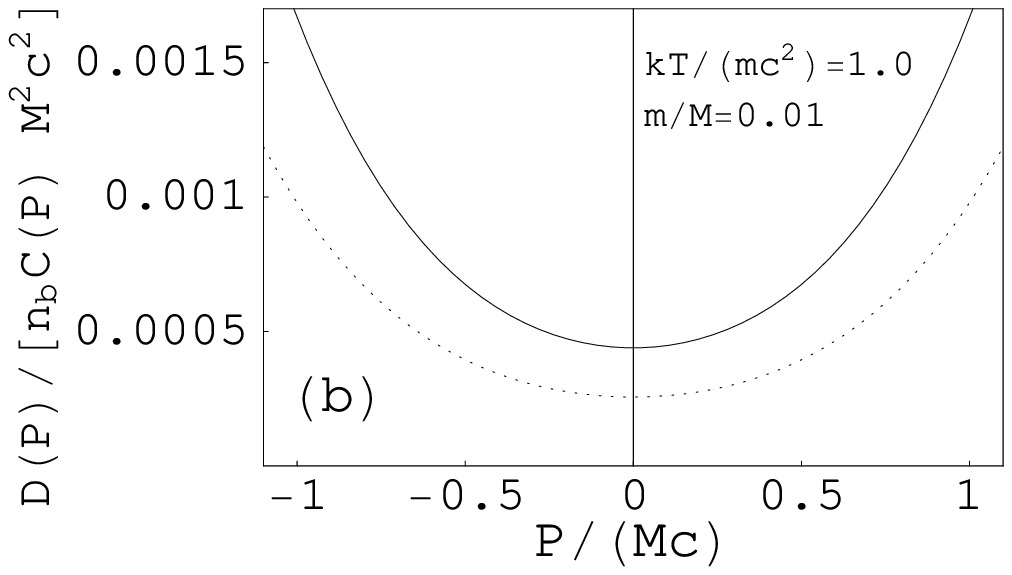 ,height=4.6cm, angle=0}
\caption{\label{fig04} Relativistic diffusion coefficients $D(P)$
calculated numerically for different heat bath distributions
$f_\mrm{b}^1(p)$ for (a) weakly  and (b) strongly relativistic heat.
The solid lines refer to a standard J\"uttner distribution with
$\eta=0$, and the dotted lines to $\eta=1$ in
Eq.~\eqref{e:Juettner-1d}. }
\end{figure}

Finally, by substituting the averaged friction coefficient
\bse\label{e:langevin_rel} 
\be\label{e:langevin_rel-a} 
\nu(P)&=&
n_\mrm{b}\;C(P)\;\lan\f{2}{1-{u}_r^2}\;
\f{\eps_r}{E+\eps_r}\ran_\mrm{b} 
\ee 
for the square bracket term in
Eq. \eqref{e:initial_1_rel}, imposing the TDL for the bath and
letting $\tau\to 0$ in Eq. \eqref{e:initial_1_rel}, we obtain the
relativistic LE 
\be \label{e:relativistic_LE} 
\dot P=-\nu(P)\, P+\chi(t), 
\ee 
where, in view of approximation~\eqref{e:product_approx}, the \emph{non-Gaussian} momentum-dependent noise force $\chi(t)\equiv\chi(P,t)$ is characterized by the mean
\be\label{e:relativistic_noise_moments} 
\mu(P)&\equiv&\lan\chi(t)\ran_\mrm{b}\notag\\
&=&n_\mrm{b}\;C(P)\; \lan \f{2}{1-u_r^2}\;
\f{E}{E+\eps_r} \,p_r\ran_\mrm{b},\qquad
\ee
and the covariance
\be
\gs(t,s)&\equiv &\notag
\lan\, \left[\chi(t)-\lan \chi(t)\ran_\mrm{b}\right]\,
\left[\chi(s)-\lan \chi(s)\ran_\mrm{b}\right]\,\ran_\mrm{b}\\
&=&
2\,D(P)\;\gd(t-s),\qquad
\ee 
with the (momentum-space) diffusion coefficient given by
\be\label{e:noise_amplitude_rel} D(P) =\f{n_\mrm{b}}{2}\,C(P)\; \lan
\left(\f{2}{1-u_r^2}\; \f{E}{E+\eps_r}\; p_r\right)^2 \ran_\mrm{b}.
\quad \ee \ese In Fig.~\ref{fig04} the ratio $D(P)/[n_\mrm{b}C(P)]$
is plotted for the same parameters as in Figs.~\ref{fig01} and
\ref{fig03}. As it is evident from the diagrams, this quantity
increases with temperature $T$ and absolute momentum $P$ of the
Brownian particle.

\section{Resume}
We  conclude the derivation of the relativistic LE with a set of
general remarks:
\par(i)
While deriving the relativistic LE~\eqref{e:langevin_rel}, we made
use of the stationarity, independence, and homogeneity of the bath
distribution~\eqref{e:rel_pdf_ansatz}; we did not, however,  rely on
the specific properties of the marginal momentum PDF. Hence, the
above results hold true for arbitrary one-particle momentum
distributions~$f_\mrm{b}^1(p)$.
\par(ii)
In order to be able to use the LEs~\eqref{e:langevin_rel} derived above, one still needs to
calculate the mean collision rate $C(P)/L$, which is determined by
Eq.~\eqref{e:rate-b}; cf. Appendix. We also emphasize once
again that the approximation~\eqref{e:product_approx}, leading to
the appearance of $C(P)$, can in principle be omitted (in the
non-relativistic as well as in the relativistic case). More precise
results for  friction coefficients and noise correlations can then
be extracted from Eq.~\eqref{a-e:mean} in the Appendix.
\par(iii)
The stochastic force $\chi(t)$ in Eq.~\eqref{e:langevin_rel} is
$\gd$-correlated (memory-free), but non-Gaussian; i.e., in order to
completely specify the stochastic process one actually has to
determine all higher order correlation functions. This is
practically unfeasible. Therefore, in numerical studies and/or
practical applications, one could use a Gaussian
approximation (GA) of Eqs.~\eqref{e:langevin_rel}, obtained in the following manner:
We rewrite Eq.~\eqref{e:relativistic_LE} equivalently as 
\bse
\be\label{e:relativistic_LE_gauss}
\dot P=-\bar{\nu}(P)\, P+\sqrt{2D(P)}\,\bar{\zeta}(t),
\ee 
where 
\be 
\bar{\nu}(P)\equiv   \nu(P)-\f{\mu(P)}{P}, \qquad
\bar{\zeta}(t)\equiv \f{\chi(t)-\mu(P)}{\sqrt{2D(P)}}.\quad
\ee
Reminiscent of standardized Gaussian white noise, the effective noise force $\bar{\zeta}(t)$ is characterized by
\be
\lan\bar{\zeta}(t)\ran=0,\qquad \lan\bar\zeta(t)\bar\zeta(s)\ran=\gd(t-s),
\ee 
\ese
but the higher moments are non-Gaussian. Accordingly, the GA is achieved by replacing $\bar{\zeta}(t)$ in Eq.~\eqref{e:relativistic_LE_gauss}  with standardized momentum-independent Gaussian white noise $\zeta(t)$. The resulting stochastic differential equation is a standard LE with multiplicative Gaussian white noise.  Hence, after having specified a discretization rule, one can easily write down  the corresponding Fokker-Planck equation as well as the corresponding stationary distribution~\cite{Kl94}. 
\par
The GA obtained this way neglects higher-order cumulants of the noise, so it cannot be expected that the \lq
truncated\rq\space LE yields exactly the same relaxation behavior and/or
the same stationary solution as the full relativistic LE~\eqref{e:langevin_rel}~\cite{Ha82,Han80}. Nevertheless, this approximation should provide useful estimates. In particular, if the stationary momentum distribution of the Brownian particle can be guessed by other arguments~\cite{DuHa06}, then the GA can be made  self-consistent with respect to this distribution by fixing a suitably generalized Einstein relation for the friction and noise coefficients. In this case it suffices to calculate, e.g., $\bar{\nu}(P)$, because the corresponding noise amplitude  $\bar{D}(P)$ is then uniquely determined by the Einstein relation (i.e., by the stationary distribution).
\par 
Due to the multiplicative noise coupling, the results obtained from the GA will also  depend on the choice of the discretization
rule~\cite{Ito44,Ito51,Fisk63,St64,St66,Kl94,Han78,Han80,HaTo82}.
Loosely speaking, this discretization dilemma is the
price that one has to pay for mapping the large number of collisions
between $t$ and $t+\tau$ onto a single instant of time. Our
experience with the non-relativistic LE (cf. remarks at the end of
Sec.~\ref{langevin-nonrel}) suggests that the "transport" or "kinetic" interpretation, corresponding to the post-point discretization rule~\cite{Han78,Han80,HaTo82,Kl94}, should be preferable in the relativistic case as well. 
\par(iv)
In principle, it should be straightforward to generalize
the above approach to higher space dimensions, by expressing
the momentum vector after the collision, $\hat{\bs P}$,
in terms of the momenta  before the collision, $\bs P$
and $\bs p$, analogous to Eq.~\eqref{e:rel_delta}.  In
the $2d$ or $3d$-case, complications may arise mostly
due to the fact that one also has take into account the
corresponding collision angles and cross-sections (e.g.,
when determining the collision rates; cf. comments at the
end of the Appendix).
\par(v)
According to our above results, the previously proposed 
\lq relativistic\rq\space LEs~\cite{DeMaRi97,Zy05,DuHa05a,DuHa05b}
should be viewed as approximations, which can be fruitful for
generating/simulating ensembles of  relativistic particles in a
simple manner. It is also evident now why these earlier approaches
were intrinsically limited. Debbasch et al. \cite{DeMaRi97} have
postulated that the relativistic stochastic force in the rest-frame 
of the bath is ordinary Gaussian white noise with a constant amplitude $D$, 
whereas we in our prior work~\cite{DuHa05a,DuHa05b} assumed 
 Gaussian noise in the (comoving) rest-frame of the Brownian particle. 
As follows from the derivation presented here, neither of these assumptions 
is accurate if one properly takes into account both the relativistic 
conservation laws and the relativistic momentum distributions of the heat bath
particles. However, if we consider suitably chosen momentum-dependent 
friction and diffusion coefficients, then the previously proposed 
\lq relativistic\rq\space LEs~\cite{DeMaRi97,Zy05,DuHa05a,DuHa05b} become equivalent 
to the Gaussian approximation of the relativistic LE~\eqref{e:langevin_rel}.

\begin{acknowledgments}
The authors would like to thank S.~Hilbert and P.~Talkner for helpful discussions.
\end{acknowledgments}

\bibliography{TD,RelTD,Papers,RelDiff,RBM,ActiveBM,RBMapplied,FokkerPlanck,BrownianMotion,RelKin,Journals,Books,PhotonsDiffusion,QBM,QFT,MathBooks,StochCalc,RelMany}


\appendix

\section{Calculation of the collision rate}
\label{a:solution}

We aim to derive an explicit expression for the expectation value
$\lan I_r(t,\tau)\ran_\mrm{b}$ in the limit $\tau\to 0$, as, e.g., required in Eqs.~\eqref{e:nu_nonrel}.
\par
By definition, the function $I_r(t,\tau)\in \{0,1\}$ indicates whether
or not the Brownian particle has collided with the heat bath particle
$r$ during the time interval~\mbox{$[t,t+\tau]$}. The positions of the
Brownian and heat bath particle at time $t$ are denoted by $X$ and $x_r$,
respectively. Ignoring the possibility of a collision, for small enough $\tau$,
the new positions at time $t+\tau$ would be given by
\be\label{a-e:indicator_0}
X'=X+V\tau,\qquad x'_r=x_r+v_r\tau,
\ee
where $V$ and $v_r$ are the velocities.
Then, the indicator function $I_r(t,\tau)$ can be explicitly represented as
\be
I_r(t,\tau)
&=&\notag
\Gt(X-x_r)\;\Gt(x'_r-X')\;\Gt(v_r-V)+\\
&&\;
\Gt(x_r-X)\;\Gt(X'-x'_r)\;\Gt(V-v_r),\qquad
\label{a-e:indicator_1}
\ee
where $\Gt(x)$ is the Heaviside-function, defined by
\be
\Gt(x)=
\begin{cases}
0,& x< 0;\\
1/2,&x=0;\\
1, & x> 0.
\end{cases}
\ee
The first (second) summand in Eq.~\eqref{a-e:indicator_1} refers
to the initial configuration, where the heat bath particle is
located at the left (right) side of the Brownian particle. Let
us list some properties of the collision indicator  $I_r(t,\tau)$.
\par
First we note that $I_r(t,\tau)$  is idempotent, i.e.,
\bse\label{a-e:properties}
\be\label{e:idempotent}
I_r^j(t,\tau)=I_r(t,\tau)
\ee
holds for $j=1,2\ldots$. Furthermore, for $\tau\to 0$, we have
\be
I_r(t,0)=0.
\ee
Accordingly, the Taylor-expansion at $\tau=0$ gives
\be
I_r(t,\tau)\simeq \left[\f{\p I_r}{\p\tau}(t,0)\right]\;\tau.
\ee
In order to determine $\lan \f{\p I_r}{\p\tau}(t,0)\ran_\mrm{b}$, we note that
\be
\f{\p}{\p\tau}\Gt(x'_r-X')\biggl|_{\tau=0}
&=&\notag
\f{\p}{\p\tau}\Gt(x_r-X+(v_r-V)\tau)\biggl|_{\tau=0}\\
&=&\notag
(v_r-V)\;\gd(x_r-X+(v_r-V)\tau)\biggl|_{\tau=0}\\
&=&\notag
(v_r-V)\;\gd(x_r-X),
\ee
and, analogously,
\be
\f{\p}{\p\tau}\Gt(X'-x'_r)\biggl|_{\tau=0}
&=&\notag
\f{\p}{\p\tau}\Gt(X-x_r+(V-v_r)\tau)\biggl|_{\tau=0}\\
&=&\notag
(V-v_r)\;\gd(X-x_r+(V-v_r)\tau)\biggl|_{\tau=0}\\
&=&\notag
(V-v_r)\;\gd(X-x_r).
\ee
Hence, we find
\be
\f{\p I_r}{\p\tau}(t,0)
&=&\notag
(v_r-V)\;\Gt(X-x_r)\;\gd(x_r-X)\;\Gt(v_r-V)+\\
&&\;\notag
(V-v_r)\;\Gt(x_r-X)\;\gd(X-x_r)\;\Gt(V-v_r),\\
&=&\notag
\Gt(0)\;(v_r-V)\;\gd(x_r-X)\;\Gt(v_r-V)+\\
&&\;\notag
\Gt(0)\;(V-v_r)\;\gd(X-x_r)\;\Gt(V-v_r),
\ee
and, with $\Gt(0)=1/2$, the useful result
\be
\f{\p I_r}{\p\tau}(t,0)&=&\notag
\f{1}{2}(v_r-V)\;\;\gd(x_r-X)\;\Gt(v_r-V)+\\
&&\;
\f{1}{2}(V-v_r)\;\gd(X-x_r)\;\Gt(V-v_r).
\qquad
\label{a-e:indicator_derivative}
\ee
\ese
\par
Now let us consider a spatially homogeneous one-particle bath distribution of the form
\be
\tilde{f}_\mrm{b}^1(x_r,v_r)=\f{1}{L}\;\tilde{f}_\mrm{b}^1(v_r)
\begin{cases}
1,&x_r\in[-L/2,L/2];\\
0,&x_r\not\in[-L/2,L/2],
\end{cases}
\ee
and some function $\tilde{G}(x_r,v_r)$ such that the expectation
value $\lan\tilde{G}(x_r,v_r)\ran_\mrm{b}$ exists. We are
interested in expectations of the form
\bse\label{a-e:mean}
\be\notag
\lan\tilde{G}(x_r,v_r)\,I_r^j(t,\tau)\ran_\mrm{b}
\overset{\eqref{e:idempotent}}{=}
\lan\tilde{G}(x_r,v_r)\,I_r(t,\tau)\ran_\mrm{b},
\ee
as required for calculating the mean value of the stochastic
force and its higher correlation functions~[e.g., compare first
line of Eq.~\eqref{e:intermediate_results}].  For small $\tau$, we may truncate
the Taylor expansion after the linear term, yielding
\be
\lan\tilde{G}(x_r,v_r)\,I_r(t,\tau)\ran_\mrm{b}\simeq
\lan\tilde{G}(x_r,v_r)\,\f{\p I_r}{\p \tau}(t,0)\ran_\mrm{b}\; \tau.\qquad
\ee
Making use of the result~\eqref{a-e:indicator_derivative}, the mean value on the rhs. is given by
\be
\lan\tilde{G}(x_r,v_r)\,\f{\p I_r}{\p \tau}(t,0)\ran_\mrm{b}
&=&\notag
\f{1}{2L}
\int_{V}^{\infty}\!\!\!\diff v_r\;(v_r-V)\;\times \\
&&\notag \qquad\quad
\tilde{G}(X,v_r)\;
\tilde{f}_\mrm{b}^1(v_r)\;+\\
&&\;\notag
\f{1}{2L}
\int^{V}_{-\infty}\!\!\!\diff v_r\; (V-v_r)\;\times \\
&&\qquad\quad\notag
\tilde{G}(X,v_r)\;
\tilde{f}_\mrm{b}^1(v_r).
\qquad\\
\ee\label{a-e:expectation_result}
\ese
In particular, by choosing $\tilde{G}(x_r,v_r)\equiv 1$, we find the collision rate
\bse\label{a-e:rate}
\be\label{a-e:rate-a}
\lim_{\tau\to 0} \f{\lan I_r(t,\tau)\ran_\mrm{b}}{\tau}
=\lan \f{\p I_r}{\p\tau}(t,0)\ran_\mrm{b}=
\f{1}{L}\;\tilde{C}(V),
\ee
where
\be
\tilde{C}(V)&\equiv&\notag
\f{1}{2}
\int_{V}^{\infty}\!\!\!\diff v_r\;(v_r-V)\;
\tilde{f}_\mrm{b}^1(v_r)+\\
&&\;\label{a-e:rate-b}
\f{1}{2}
\int^{V}_{-\infty}\!\!\!\diff v_r\; (V-v_r)\;
\tilde{f}_\mrm{b}^1(v_r).
\ee
\ese
The following comments are in order:
\par(i)
The above derivation is valid for both non-relativistic and relativistic
heat bath distributions $\tilde{f}^1_\mrm{b}(v_r)$. Upon
identifying $C(P)\equiv \tilde{C}(V(P))$, where $P$ is the
momentum of the Brownian particle, we obtain the rigorous
justification for Eq.~\eqref{e:Stokes-a}. However, in the
non-relativistic case we have $V=P/M$, whereas in the
relativistic case $V=P/(M^2+P^2)^{1/2}$. Additionally,
we note that the support interval of the relativistic
velocity distribution $\tilde{f}^1_\mrm{b}(v_r)$ is given by $[-c,c]$,
which determines the effective upper and lower integral boundaries in Eq.~\eqref{a-e:rate-b}.
\par (ii)
Given a certain bath distribution $\tilde{f}^1_\mrm{b}(v_r)$, the exact
result~\eqref{a-e:mean} allows for evaluating the quality of the
product approximations~\eqref{e:Stokes-b} and \eqref{e:product_approx}, respectively.
\par (iii)
The Stokes approximation corresponds to setting $V=0$ in Eq.~\eqref{a-e:rate-b}, yielding
\be
\tilde{C}(0)=
\f{1}{2}
\int_{-\infty}^{\infty}\diff v_r\;|v_r|\;
\tilde{f}_\mrm{b}^1(v_r).
\ee
This shows that the Stokes approximation is useful for slow Brownian particles,
but inappropriate at high velocities.
\par (iv)
It is in principle possible to apply the same procedure to higher space
dimensions, but then the expression~\eqref{a-e:indicator_1} for the indicator
unction has to be modified accordingly (e.g., by taking into account the
geometric shape of the Brownian particle). As a consequence, analytic
calculations will become much more difficult.
\end{document}